\documentclass[11pt,twoside,dvips]{article}
\usepackage{verbatim}
\usepackage{amssymb}
\usepackage{amsfonts}
\usepackage{amsmath}
\usepackage{pifont}
\usepackage[spanish,english]{babel}
\usepackage[pdftex]{graphicx}
\usepackage{epsfig}
\usepackage{bm}
\usepackage{longtable}
\usepackage{fancyhdr}

\begin{document}

\fancypagestyle{plain}{%
\fancyhf{}%
\fancyhead[LO, RE]{XXXVIII International Symposium on Physics in Collision, \\ Bogot\'a, Colombia,  
11-15 September 2018}}

\fancyhead{}%
\fancyhead[LO, RE]{XXXVIII International Symposium on Physics in Collision, \\ Bogot\'a, Colombia,  
11-15 September 2018}

\title{Electroweak phase transition in the scotogenic model}
\author{Alejandro Correa L\'opez$\thanks{e-mail: alejandro.correal@udea.edu.co}$, \\ Instituto de 
F\'isica, Universidad de 
Antioquia,\\  Calle 70 \# 52-21, Apartado A\'ereo 1226, Medell\'in, Colombia}
\date{}
\maketitle

\begin{abstract}
In this work we study the behavior of the effective scalar potential at zero and finite temperature 
of the scotogenic model. In particular, we analyze the impact of the Yukawa couplings associated to 
the right--handed neutrinos on the electroweak phase transition. 
\end{abstract}
%

\section{Introduction}

The need for physics beyond the Standard Model (SM) arises from experimental evidence of at least 
three phenomena with no support within the SM framework: neutrino masses, dark matter (DM) and the 
matter--antimatter asymmetry. Along this lines, Ernest Ma \cite{Ma2006} proposed the scotogenic 
model, that radiatively generates neutrino masses and has a viable DM candidate. On the other hand, 
the matter--antimatter asymmetry can be explained from two general approaches: baryogenesis via 
leptogenesis \cite{Strumia2006} and electroweak baryogenesis (EWB) \cite{Morrissey2012}. Sakharov 
\cite{Sakharov:1967dj} set three conditions for any model that attempts to explain the baryon 
asymmetry: {\it i)} baryon number violation, {\it ii)} CP violation and {\it iii)} departure from 
thermal equilibrium. In this work we will only focus on the third condition. Within EWB, the 
departure from equilibrium is achieved through a strong first--order electroweak phase 
transition\footnote{This means an energetic barrier between two minimum at different values of the 
vacuum expectation value ({\it vev}).} (EWPT) and the baryon number preservation condition (BNPC) of 
the effective scalar potential. Thus, the aim of this work is to show how is the behaviour of the 
scalar potential of the scotogenic model when quantum and thermal corrections are taken into 
account; in particular, when the Yukawa couplings leading to neutrino masses are non negligible.


\section{The model and thermal corrections}
\label{model}
The scotogenic model is an extension of the SM with a second Higgs doublet $\Phi$, three 
right-handed neutrinos, $N_i$ ($i=1,2,3$), and a $Z_2$ symmetry such that the SM fields 
are even and the new fields are odd. The right--handed neutrino interactions with SM particles are 
given by:
\begin{eqnarray}
\mathcal{L}_{int} = -h_{ij} \overline{N_R^i} \tilde{\Phi}^{\dagger} \ell^j_L  + \text{h.c.} 
\quad \text{($\tilde{\Phi}=i\sigma_2 \Phi^*$)}\label{int-lag}.
\end{eqnarray}
On the other side, the scalar interactions of the model can be expressed via the scalar potential
\cite{Blinov:2015vma}: 
\begin{eqnarray}
 V_0 &= &\mu^2_1|H|^2 + \mu^2_2|\Phi|^2 + \lambda_1 |H|^4 + \lambda_2 |\Phi|^4  +
\lambda_3 |H|^2|\Phi|^2 +\lambda_4 |H^{\dagger}\Phi|^2  \nonumber \\
&+& \frac{\lambda_5}{2}\left[(H^{\dagger}\Phi)^2 + \text{ h.c.}\right]\label{scalar_pot-SCT},
\end{eqnarray}
where the doublets are
\begin{equation}
 \label{dobletes-SCT}
H = \begin{pmatrix} G^{\pm} \\ \frac{1}{\sqrt{2}}(v+h+iG^0) 
\end{pmatrix}, \quad
\Phi = \begin{pmatrix} H^{\pm} \\ \frac{1}{\sqrt{2}}(\phi + H^0+iA^0) 
\end{pmatrix},
\end{equation}
being $v$ and $\phi$ the {vev} of $H$ and $\Phi$, respectively, and $H^0$ is the DM candidate (for 
this work). In the most DM models, the $Z_2$ symmetry remains unbroken to ensure stability of the 
DM 
candidate. In this work, we will relax that condition and allow $Z_2$ can develop a {\it vev}.


\subsection*{The effective potential and thermal corrections}
The thermal corrections are included through the effective scalar potential:
\begin{equation}
\label{eff-pot-SCT}
 V_{eff}(v,\phi,T) = V_0 + V_1(v,\phi) + V_T(v,\phi, T), 
\end{equation}
where $V_0$ is the tree level scalar potential, $V_1(v,\phi)$ is the one loop correction to 
(\ref{scalar_pot-SCT}) at zero--temperature (Coleman--Weinberg potential) and  $V_T(v,\phi, T)$ 
represents the thermal and ring contributions \cite{Espinosa2007}:
\begin{equation}
 \label{cw_pot-SCT}
 V_1(v,\phi) = \sum_{i\,={\text{ all fields}}}\frac{n_i m_i^4(v,\phi)}{64\pi^2}\left[ 
\ln{\frac{m^2_i(v,\phi)}{Q^2}} - 
c_i\right],
\end{equation}
\begin{eqnarray}
 \label{t-contri-SCT}
 V_T(v,\phi, T) &=& \frac{T^4}{2\pi^2}\left\{\sum_{B} n_i 
J_B\left[\frac{m_i^2(v,\phi)}{T^2}\right]+\sum_{F} 
 n_i J_F\left[\frac{m_i^2(v,\phi)}{T^2}\right]\right\} \nonumber \\
 &+& \sum_{i\,={\text{ all fields}}} 
\frac{T}{12\pi}\left(\frac{1+\varepsilon}{2}\right)n_i\left\{\left[m_i^2(v,\phi)\right]^3 
-\left[m_i^2(v,\phi,T)\right]^3\right\},
\end{eqnarray}
where $B$ stands for scalars and bosons, and $F$ for fermions. The thermal functions are 
expressed as:
\begin{equation}
 J_B[y^2]=\int_0^{\infty} dx\
x^2\log\left[1-e^{-\sqrt{x^2+y^2}}\right],
\quad J_F[y^2]=\int_0^{\infty} dx\
x^2\log\left[1+e^{-\sqrt{x^2+y^2}}\right].
\end{equation}
In this case, $n_i$ are \cite{Blinov:2015vma} $n_Z=3$, $n_W=6$, $n_t=-12$, $n_h=1$, $n_{G^0}=1$, 
$n_{G^{\pm}}=2$, $n_{H^{\pm}}=2$, $n_{H^0}=1$, and $n_{A^0}=1$. The $c_i$ coefficients 
are established by the nature of the fields: $3/2$ for scalars and fermions and  $5/6$ for gauge 
bosons. The factor $\varepsilon$ is $1$ for bosons and $-1$ for fermions\footnote{Note that for 
fermions the last term of (\ref{t-contri-SCT}) vanishes, thus there is not ring contribution.}, and 
$Q$ is the renormalization scale.
\subsection*{Masses of the model}
The field dependent masses, $m_i$, are determined by a diagonalization process of the following 
matrices
\cite{Blinov:2015vma}:
\begin{eqnarray}
M^2_h  = \begin{pmatrix}
\mu_1^2 + 3\lambda_1 v^2+ \lambda_L \phi^2 & 2\lambda_L \phi v \\
 2\lambda_L \phi v & \mu_2^2 + 3\lambda_2 \phi^2 + \lambda_L v^2
\end{pmatrix} \label{eq:CPEvenMass}, \quad
M^2_A  = \begin{pmatrix}
\mu_1^2 + \lambda_1 v^2+ \lambda_S \phi^2 & \lambda_5 \phi v \\
 \lambda_5 \phi v & \mu_2^2 + \lambda_2 \phi^2 + \lambda_S v^2
\end{pmatrix}, \label{eq:CPOddMass}\nonumber
\end{eqnarray}
\begin{eqnarray}
M^2_\pm =\begin{pmatrix}
\mu_1^2 + \lambda_1 v^2+ \frac{1}{2}\lambda_3 \phi^2 & \frac{1}{2}(\lambda_5+\lambda_4) \phi v \\
\frac{1}{2}(\lambda_5+\lambda_4) \phi v & \mu_2^2 + \lambda_2 \phi^2 + \frac{1}{2}\lambda_3 v^2
\end{pmatrix},
\label{eq:ChargedMass}
\end{eqnarray}
with $\lambda_L = (\lambda_3+\lambda_4+\lambda_5)/2$ and $\lambda_S = 
(\lambda_3+\lambda_4-\lambda_5)/2$. For the thermal corrections, the masses resulting from the 
diagonalization process of (\ref{eq:CPEvenMass}) contain the $\mu_i$ parameters. Thus, 
the thermal corrections require that $\mu^2_i \longrightarrow \mu^2_i + c_iT^2$ 
\cite{Blinov:2015vma,Merle2015}, where
\begin{equation}
\label{c1t}
 c_1 = \frac{1}{8} g^2 + \frac{1}{16}(g^2 + g^{\prime 2}) + \frac{1}{2}\lambda_1 
+ \frac{1}{12}(\lambda_L + \lambda_S)+ \frac{1}{12}\lambda_3
+ \frac{1}{4} y_t^2, \quad \text{(SM fields)}
\end{equation}
\begin{equation}
\label{c2t}
 c_2 = \frac{1}{8} g^2 + \frac{1}{16}(g^2 + g^{\prime 2}) + \frac{1}{2}\lambda_2 
+ \frac{1}{12}(\lambda_L + \lambda_S+ \lambda_3)+\frac{9}{12}h^2_{\nu}. \quad \text{(new fields)}
\end{equation}
being $g$ and $g'$ the electroweak couplings, $y_t$ the Yukawa top coupling and $h_{\nu}$ are given 
by
\begin{equation}
 \label{yukas}
 h_{\nu} = \frac{1}{3} \sqrt{\sum_{i,j}|h_{ij}|^2}.
\end{equation}
Note that the $h_{\nu}$ couplings contribute to the thermal masses of the new fields (see equation 
(\ref{c2t})) in opposition to the inert double model (IDM) case \cite{Blinov:2015vma}. For gauge 
bosons, only longitudinal components get thermal masses 
given by 
the eigenvalues of
\begin{equation}
\label{gauge-matrix}
\begin{pmatrix}
\frac{1}{4}(v^2 + \phi^2)g^2 +2 g^2 T^2  & -\frac{1}{4}(v^2 + \phi^2)g g^\prime \\
-\frac{1}{4}(v^2 + \phi^2)g g^\prime & \frac{1}{4}(v^2 + \phi^2)g^{\prime 2}+2 g^{\prime 2} T^2 
\end{pmatrix},
\end{equation}
while for $W$ the contribution is $m^2_{W_l} = m^2_{W_l}(v,\phi) + 2g^2T^2.$


\section{Numerical analysis and results}
\label{results}
For numerical analysis the masses and couplings are fixed as follows: $m_{H^0} = 66$ GeV (low DM 
mass regime), $m_{H^{\pm}}$ and $m_{A^0}$ greater than $200$ GeV (with $m_{H^{\pm}}=m_{A^0}$), 
$\lambda_L = \lambda_2 = 0.01$ and $h_{\nu} = 0.31,1.0,1.73$. The scale $Q$ is fixed at electroweak 
symmetry breaking scale, that is $Q= 246.22$ GeV. 
\subsection*{Limits for $m_{H^\pm},m_{A^0}$ masses}
At zero temperature, the effective scalar potential of the model (\ref{eff-pot-SCT}) must exhibit 
the usual behaviour of the SM, i.e., the global minimum of the system must be associated with the 
electroweak symmetry breaking. This imposes limits on the masses of $m_{H^\pm}$ and $m_{A^0}$.
 \begin{figure}[h!]
\centering
\includegraphics[width=.45\linewidth]{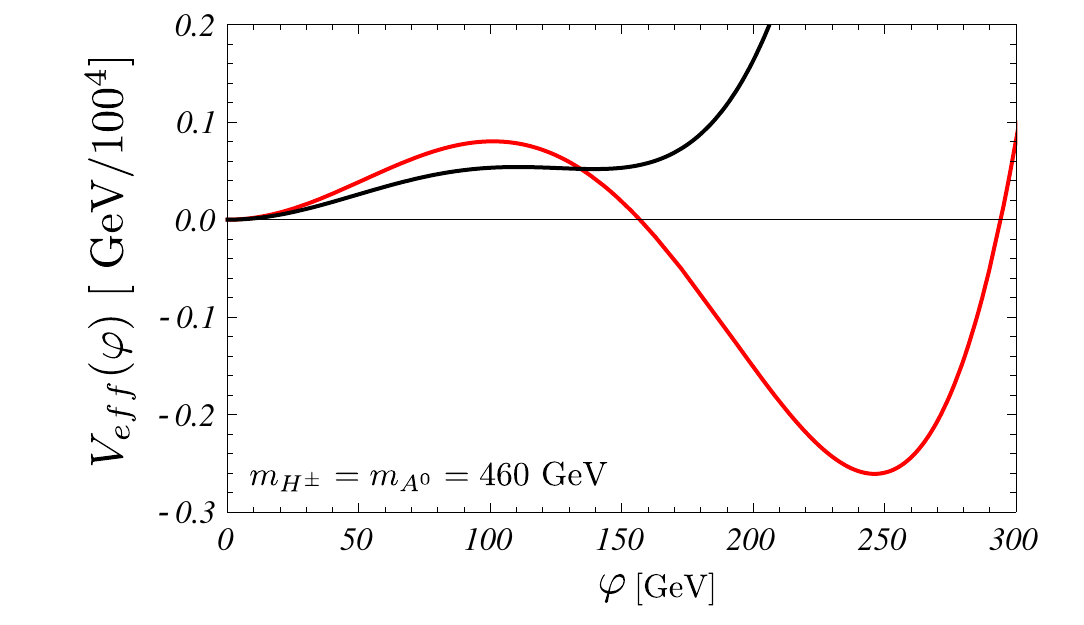}
\includegraphics[width=.45\linewidth]{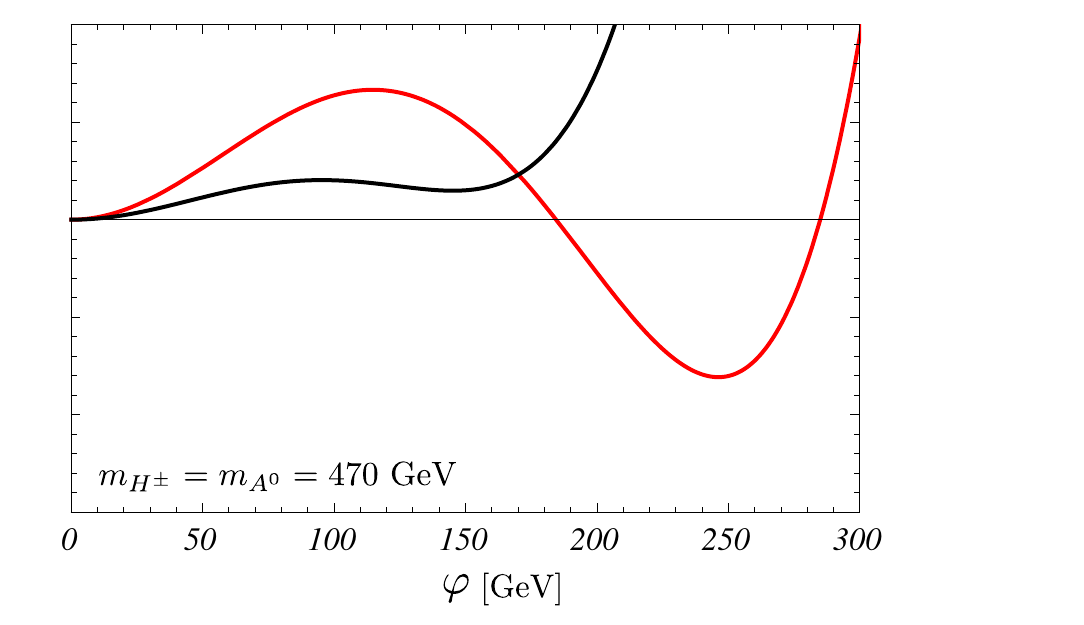}
\\[\baselineskip]
\includegraphics[width=.45\linewidth]{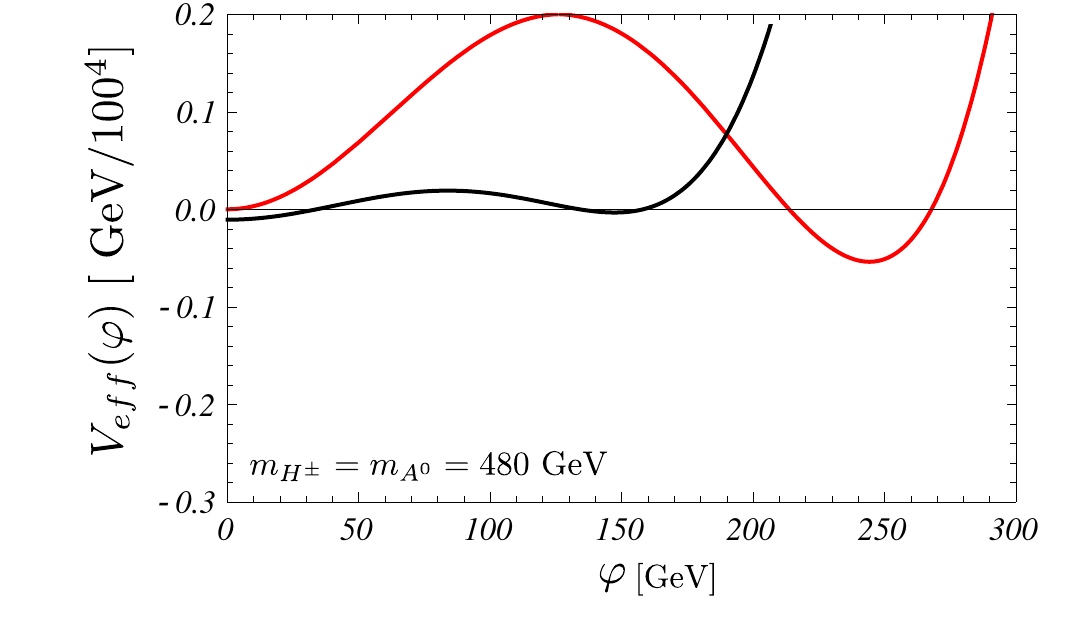}
\includegraphics[width=.45\linewidth]{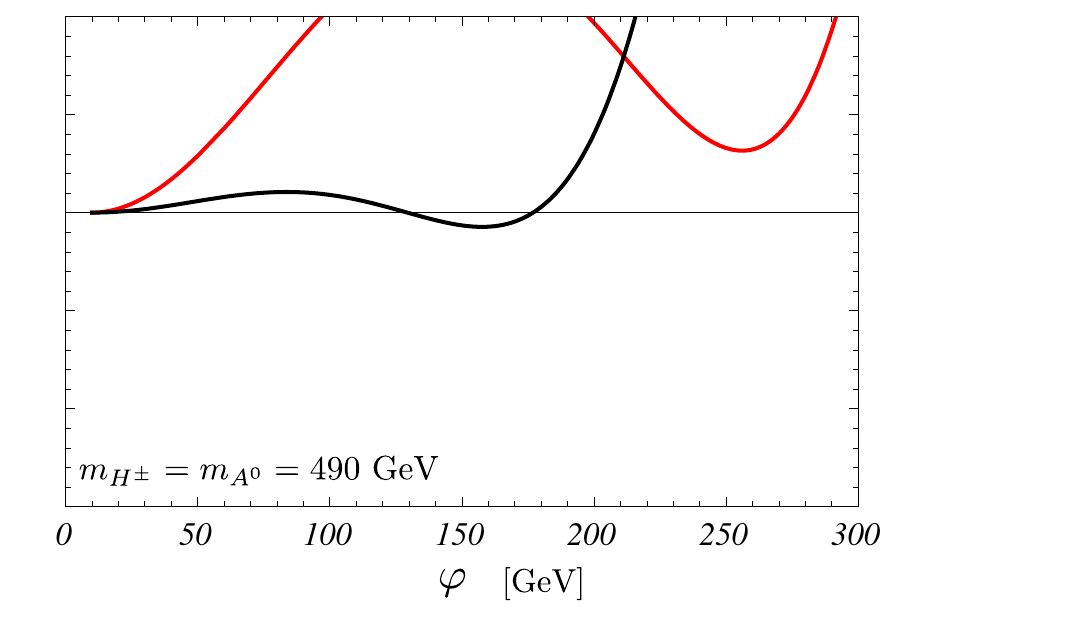}
 \caption{Effective potential behaviour at $T=0$ GeV when $m_{H^0} = 66$ GeV and different values 
of $m_{H^\pm}=m_{A^0}$. The red line indicates $V_{eff}(\varphi) = V_{eff}(v,0,0)$ and black line 
indicates $V_{eff}(\varphi) = V_{eff}(0,\phi,0)$.}
 \label{H066-Hpmdiff}
 \end{figure}
As we can see in Figure \ref{H066-Hpmdiff}, the behaviour of $V_{eff}(v,\phi,0)$ is related with 
the values of $m_{H^\pm}$ and $m_{A^0}$. For $m_{H^\pm}=m_{A^0}=460$ GeV, the effective potential 
$V_{eff}(v,0,0)$ (red line in Figure \ref{H066-Hpmdiff}) have a deeper global minimum respect 
$V_{eff}(0,\phi,0)$ (black line in Figure \ref{H066-Hpmdiff}). Thus the phenomenology of the model 
is associated to SM phenomenology. As the value of the masses increases, this global minimum begins 
to rise while $V_{eff}(0,\phi,0)$ begins to sink. However, when $m_{H^\pm}= m_{A^0} = 
490$ GeV, the global minimum of the effective potential is lost with the $Z_2$ symmetry 
breaking, which cannot be possible because the SM phenomenology. So, in order to 
preserve the electroweak symmetry breaking, the masses of $m_{H^\pm}$ and 
$m_{A^0}$ GeV 
are set at 460 GeV hereafter. 
\subsection*{EWPT}
\begin{figure}[h!]
\centering
\includegraphics[width=.47\linewidth]{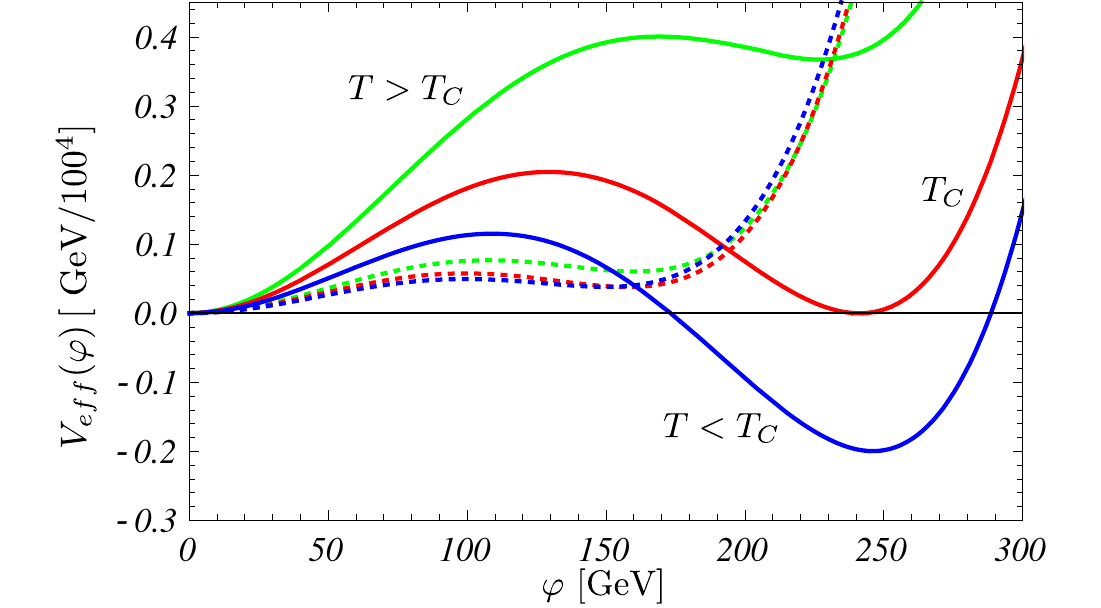}
\includegraphics[width=.47\linewidth]{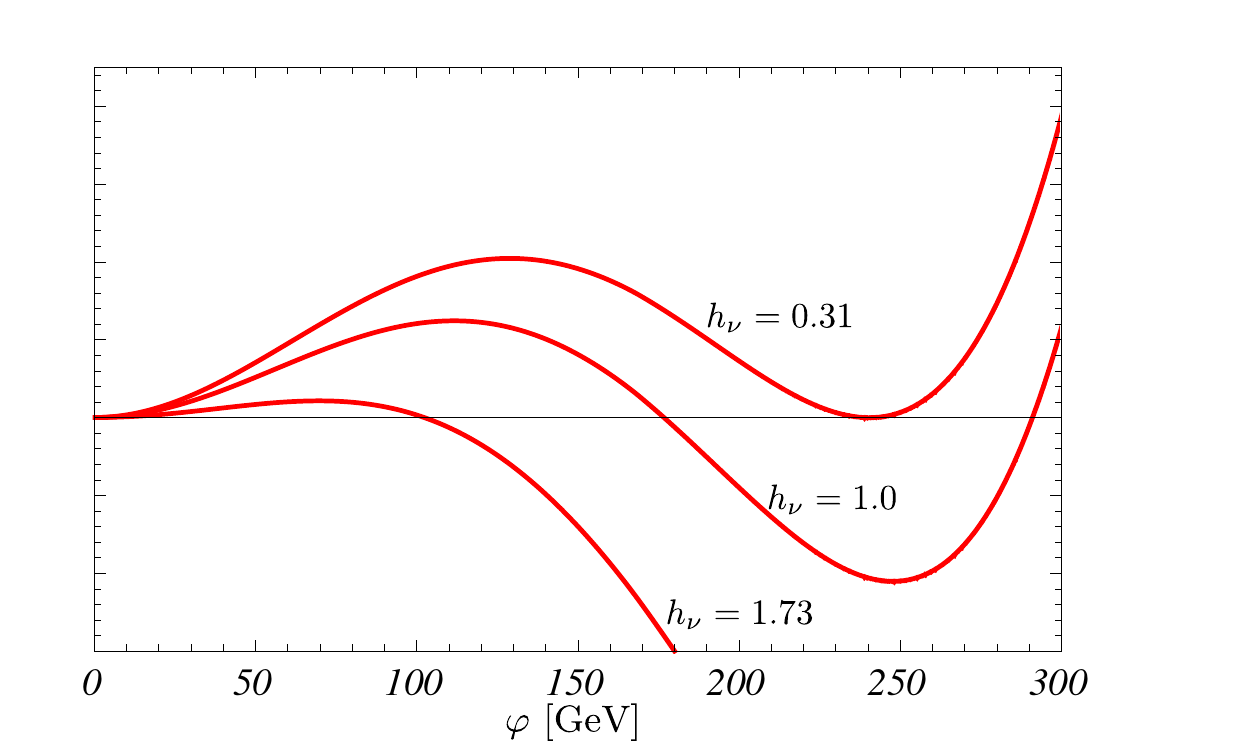}
 \caption{Left: effective potential behaviour for different temperature values and $h_{\nu}=0.31$. 
The solid lines indicate $V_{eff}(\varphi) = V_{eff}(v,0,T)$ and the dashed lines indicate 
$V_{eff}(\varphi) = V_{eff}(0,\phi,T)$. Right: effective potential behaviour at $T=90.8$ 
GeV, $m_{H^0}= 66$ GeV and $m_{H^\pm}= m_{A^0} = 460$ GeV for different values of $h_{\nu}$.}
 \label{temp-evolution}
 \end{figure}

One of the main conditions needed to have a strong first-order EWPT is the BNPC, which requires that 
$ R \equiv v_C/T_C >1,$  \cite{Blinov:2015vma}, where the critical temperature $T_C$ is the 
temperature at which the potential (\ref{eff-pot-SCT}) has two degenerate minima (for different 
values of the SM {\it vev}), and $v_C$ is the non-zero {\it vev} of the broken phase at the critical 
temperature. For $m_{H^\pm}= m_{A^0} = 460$ GeV and $h_{\nu}=0.31$, the effective potential exhibits 
a strong first--order EWPT at a critical temperature of $T_C= 90.8$ GeV when $v_C = 242.92$ GeV 
(and therefore $R=2.67$), as it is shown in left panel of Figure \ref{temp-evolution} (red solid 
line).
The interesting thing about this benchmark point is that the $Z_2$ symmetry is not affected by the 
thermal corrections and never develops a second minimum deeper than the SM one. Therefore, the 
DM candidate is still stable.
\subsection*{Impact of the Yukawas on the effective potential}
Another interesting aspect is the dependence of the effective potential with the Yukawa couplings 
$h_{\nu}$. From the right panel of Figure \ref{temp-evolution}, it is clear that the strong 
first--oder EWPT only occurs when the Yukawa couplings are set to $h_{\nu}= 0.31$. Moreover, we 
found that for $h_{\nu}=1.0$ the EWPT is still strong and first order but with a higher critical 
temperature $T'_C=129.8$ GeV at $v'_C= 244.3$ GeV ($R'=1.89$). Whereas for $h_{\nu}= 1.73$, the 
behaviour of the scalar effective potential changes and therefore it is not possible to have strong 
first--order EWPT. 

\section{Conclusions}
\label{conclude}

The scotogenic model allows us to establish the minimum conditions to explain two of the phenomena 
that the SM is not able to explain given the current data. However, may have one of the 
requirements to address a third phenomena: the baryon asymmetry. In this work we have studied the 
EWPT and BNPC in the scotogenic model through the effective scalar potential. Taking into account 
only the Coleman--Weinberg potential ($T=0$), we can set a limit of 
$m_{H^\pm}=m_{A^0} = 460$ GeV for the validity of SM. When the thermal corrections of the effective 
potential and masses of the fields are considered, the Yukawa coupling values have a strong impact 
on the behaviour of the effective potential. When the DM candidate has a mass of $m_{H^0}= 66$ GeV, 
$m_{H^\pm}= m_{A^0} = 460$, and $h_{\nu}= 0.31$, the effective potential exhibits a strong 
first--order EWPT and the BNPC is satisfied. However, for the same benchmark point and larger 
values of $h_{\nu}$, the strong first--order EWPT and BNPC are only present when $h_{\nu}= 1.0$ for 
a critical temperature such that $T'_C>T_C$ (and $R'< R$). Consequently, a strong first--order EWPT 
will be present only if $h_{\nu}\leqslant 1.0$ for the scotogenic model.



\end{document}